\documentclass[aps,superscriptaddress,twocolumn,twoside,floatfix,prb,a4paper]{revtex4-2}
\usepackage{times}
\usepackage{epsfig}
\usepackage{amsfonts}
\usepackage{amsmath}
\usepackage{amssymb,amsthm}
\usepackage{xcolor,colortbl}
\usepackage{multirow}
\usepackage{braket}
\usepackage{latexsym}
\usepackage{amsfonts}
\usepackage{mathrsfs}
\usepackage{natbib}
\usepackage{verbatim}
\usepackage{gensymb}
\usepackage{caption}
\usepackage{ragged2e}
\DeclareCaptionJustification{justified}{\justifying}
\usepackage{blkarray}
 \usepackage{graphicx}
 \usepackage{enumerate}
\usepackage[shortlabels]{enumitem}
\usepackage{ulem}
\usepackage{chemformula}
\usepackage{soul}

\usepackage[colorlinks=true,linkcolor=red,citecolor=blue,urlcolor=blue]{hyperref}
\allowdisplaybreaks

\begin{document}

\author{ Pradeepa  H L}
\email{hl.pradeepa@acads.iiserpune.ac.in}
\affiliation{Department of Physics, Indian Institute of Science  Education and Research(IISER), Pune 411008, India}
\author{ Sagnik Chatterjee}
\affiliation{Department of Physics, Indian Institute of Science  Education and Research(IISER), Pune 411008, India}
\author{Sayantan Patra}
\affiliation{Department of Physics, Indian Institute of Science  Education and Research(IISER), Pune 411008, India}
\author{ Swapneswar Bisoi}
\affiliation{Department of Physics, Indian Institute of Science  Education and Research(IISER), Pune 411008, India}
\author{Saqlain Mushtaq}
\affiliation{Department of Physics, Indian Institute of Science  Education and Research(IISER), Pune 411008, India}
\author{Hardeep}
\affiliation{Department of Physics, Indian Institute of Science, Bangalore 560012, India}
\author{Akshay Singh}
\affiliation{Department of Physics, Indian Institute of Science, Bangalore 560012, India}
\author{Ashish Arora}
\affiliation{Department of Physics, Indian Institute of Science  Education and Research(IISER), Pune 411008, India}
\author{Atikur Rahman}
\email{atikur@iiserpune.ac.in}
\affiliation{Department of Physics, Indian Institute of Science  Education and Research(IISER), Pune 411008, India}

\title{Exciton-Selective Phonon Coupling in a Lead Halide Perovskite}


\begin{abstract}
Exciton–phonon interactions govern the optical response of semiconductors, yet disentangling multiple coupling channels in lead halide perovskites remains challenging. We investigate \ch{CsPbBr3} microcrystals using photoluminescence, Raman and reflectance spectroscopy at low temperature, revealing the simultaneous presence of high-energy and Rashba excitons, each accompanied by distinct phonon replica series. High-energy exciton replicas are uniquely spaced by approximately 9 meV, whereas Rashba exciton replicas exhibit a characteristic approximately 6 meV spacing,  indicating the specificity of the exciton-phonon coupling. Unsupervised machine learning applied to a large low-temperature photoluminescence dataset reveals these replica features are prevalent.  With increasing temperature, replica features broaden and merge, evolving into a dominant longitudinal optical phonon coupling regime at room temperature. This work establishes direct spectroscopic evidence for concurrent, exciton-specific phonon coupling within a single material, offering new pathways to engineer light–matter interactions for optoelectronic  and phonon-photon-based quantum device applications.
\end{abstract}

\maketitle

Excitons, bound electron-hole pairs, are key quasiparticles shaping the optical properties of semiconductors and their interaction with phonons profoundly influences spectral characteristics and energy dissipation, making the elucidation of these mechanisms crucial for the rational design of next-generation materials \cite{baranowski2020excitons,yamada2022electron}. Due to their soft crystal lattices, lead-halide perovskites (\ch{APbX3}; A = Cs, MA, etc.; X=Cl, Br, I) exhibit strong exciton-phonon coupling \cite{iaru2021frohlich, cho2021luminescence}. In addition, they possess extraordinary optical properties, including a high photoluminescence (PL) quantum yield, strong light absorption, long carrier lifetimes, and a widely tunable bandgap. Together, these remarkable features make them both fundamentally interesting and highly promising for optoelectronic and energy applications \cite{akkerman2018genesis, poonia2024emerging}. 

Precise control of exciton-phonon interactions in these materials is pivotal for advancing optoelectronics, from solar cells to quantum computing platforms\cite{akkerman2018genesis,becker2018bright}. Amidst these materials, \ch{CsPbBr3} is highly studied for its intriguing optical response\cite{baranowski2020excitons,akkerman2018genesis,becker2018bright, kharintsev2024extreme,wu2018perovskite,wei2016temperature}, but its fundamental properties, especially the nature of excitonic states and their coupling with phonons, remain debatable \cite{baranowski2020excitons,poonia2024emerging,fu2023carriers,baranowski2019giant,yamada2022electron, Abbas_2025}. The existence and precise nature of Rashba excitons are of particular interest\cite{steele2019role,wu2019indirect,kepenekian2017rashba,becker2018bright,isarov2017rashba,dendebera2020time,dendebera2022temperature}, with origins attributed to polar fluctuations and lattice distortions\cite{wu2019indirect,yaffe2017local,yao2025crystal}. Identifying the spectroscopic fingerprints of these debated excitons and their coupling is paramount. Despite extensive effort, the direct and simultaneous observation of multiple excitonic species, each with a distinct phonon replica series, is scarce\cite{yao2025crystal, cho2022exciton,isarov2017rashba,guo2019dynamic,iaru2017strong,tamarat2023universal,dendebera2022temperature,bulyk2023pressure}. Specifically, the interplay between high-energy excitons and Rashba excitons and their coupling to different phonon modes remains unresolved in a single material system.

In this article, we intend to provide direct evidence, not only of the debated Rashba exciton in \ch{CsPbBr3}, but also for its intriguing and distinct phonon coupling behaviour, offering a critical contribution to this long-standing discussion.  Using synergistic low temperature PL, reflectance, and Raman spectroscopy on \ch{CsPbBr3} microcrystals, we reveal the coexistence of high-energy and Rashba excitons. Crucially, each exciton exhibits an equispaced phonon replica series attributed to coupling to distinct phonon modes an uncommon observation in microcrystals, where coupling typically diminishes\cite{zhu2024quantifying,amara2023spectral,cho2022exciton}. Their temperature evolution uncovers a crossover to a dominant high-energy longitudinal optical (LO) phonon coupling regime at room temperature. We statistically confirmed the frequent occurrence of these replica features using the k-means clustering algorithm on 260 PL spectra. These findings offer fundamental understanding to control exciton-phonon dynamics for advanced opto-electronics and phonon-photon based quantum devices\cite{Ripin_2023}.

\begin{figure}[t]
 \centering
 \includegraphics[width=1\linewidth]{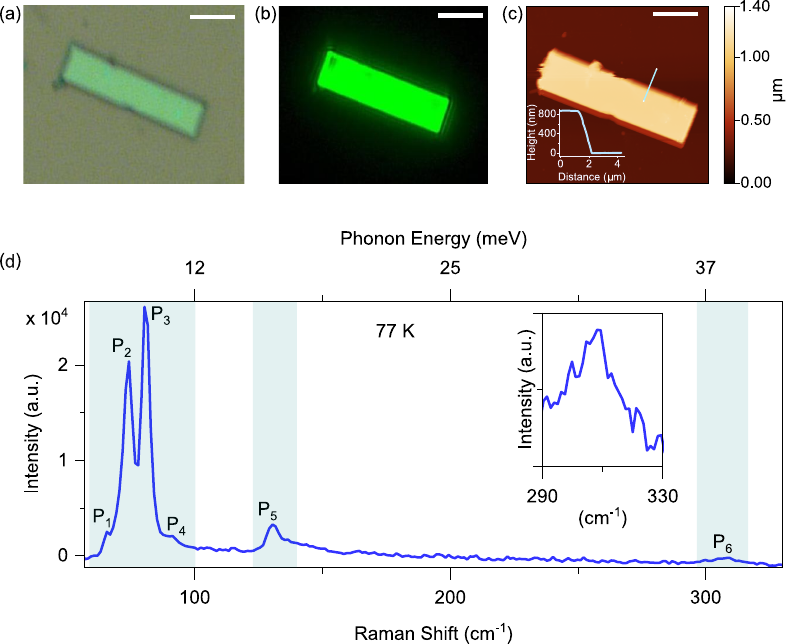}
 \caption{ \label{fig1}  Morphological and optical characterization of \ch{CsPbBr3} sample. (a) Optical image and (b) Wide-field PL image of the \ch{CsPbBr3} sample, demonstrating strong emission. (c) Atomic force microscopy (AFM) image and corresponding height profile(inset), which shows a sample height of $\sim $0.83 $\mu\rm{m}$. (Scale bar: 5 $ \mu $m.) (e) Raman spectra at 77 K, exhibiting phonon peaks at  8 meV (P$_1$), 9 meV (P$_2$), 10 (P$_3$) and 11 meV (P$_4$), with an additional weaker peaks at 16 meV (P$_5$)  and 38 meV.  Inset is a magnified view of the peak P$_6$.}
 \end{figure}

We synthesized \ch{CsPbBr3} perovskite samples using a new near-room-temperature solvothermal method . An alcohol-based solvent enables controlled growth of large-area, highly crystalline structures under mild conditions\cite{anilkumar2024near}(see SI). Fig.~\ref{fig1} presents an overview of the morphological and optical characteristics of the \ch{CsPbBr3} sample. Optical and wide-field PL images at room temperature show strong emission with 405 nm excitation (Fig.~\ref{fig1} (a) and (b)). Atomic force microscopy (AFM) shows the crystal height ranges from 200 nm to 900 nm  Fig.~\ref{fig1} (c) (SI S2). The low-temperature (77 K) Raman spectrum  exhibits characteristic \ch{CsPbBr3} single crystal peaks (Fig.~\ref{fig1} (d)). Strong phonon peaks are observed at 9 meV  (P$_2$) and 10 meV (P$_3$), attributed to octahedra distortions\cite{hoffman2023understanding}. A less intense peak at 16 meV relates to Pb-Br bond stretching\cite{Abbas_2025}, and a longitudinal optical (LO) phonon peak is observed at 38 meV\cite{iaru2021frohlich,bataev2024electron}.

\begin{figure*}[t!]
  	\includegraphics[width=0.7\linewidth]{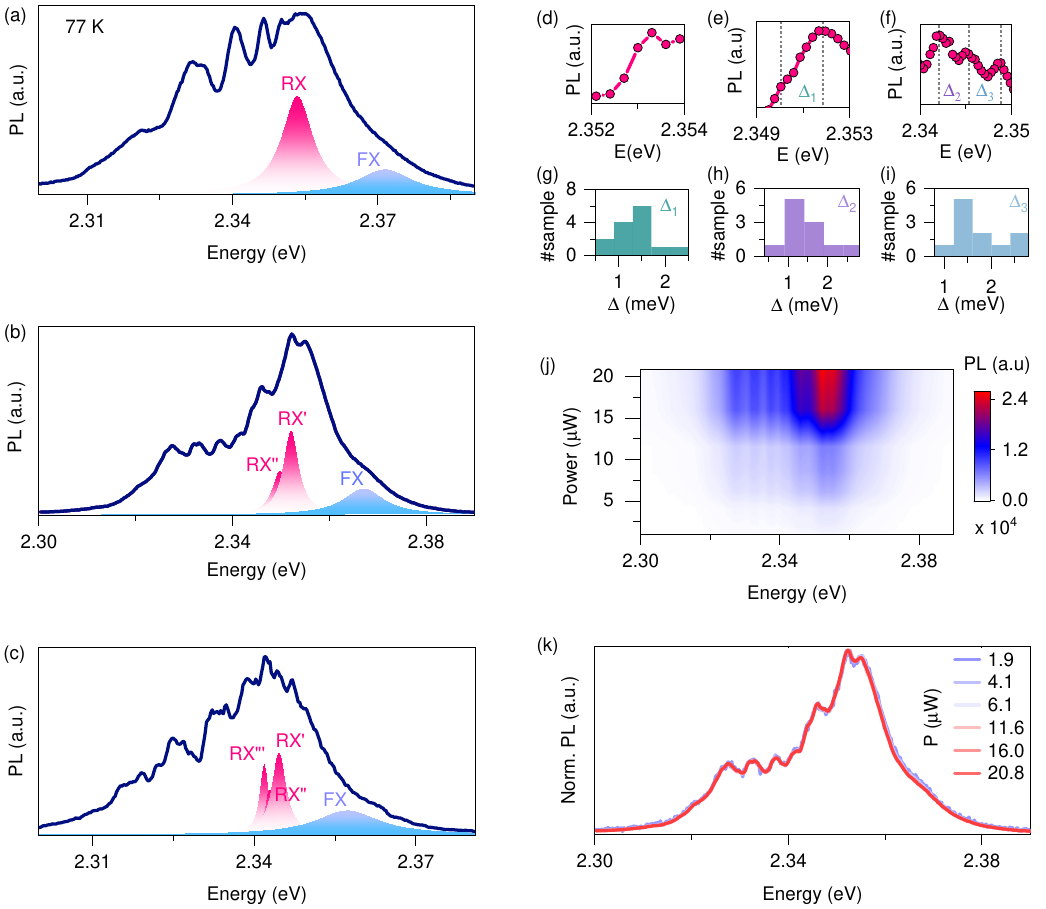}
  	\caption{ \label{fig2} Low-temperature PL and power dependence.  PL spectra at  77 K, multiple distinct emission peaks are observed, including the high energy first exciton (FX)   and Rashba exciton (RX). (a) Shows the PL of the sample  which exhibit single Rashba emission. (b) PL of the sample  with double Rashba emission (RX$'$ and  RX$''$), which appear as a peak and a hump in the spectra.  (c) shows the PL of the sample  where Rashba peak consists of three peaks (RX$'$, RX$''$ and RX$'''$ ). (d-e) The single, double(with energy separation $\Delta_1$) and triple(with energy separations $\Delta_2$ and $\Delta_3$)  Rashba emission respectively in different samples. (g-i)   $\Delta_1$, $\Delta_2$, and $\Delta_3$ distribution in different number of samples. (j) Colour map of PL intensity as a function of excitation power, showing no significant energy shift with varying power. (k) Normalised PL spectra at different excitation powers showing the low energy peaks scale like the highest intensity peak.}  
  \end{figure*}

At 77 K, upon excitation, we observed multiple clearly resolved emission peaks (Figure \ref{fig2} a-c), which can be attributed to different excitonic species with various recombination pathways. Interestingly, we observe that the first peak (FX)  at the highest energy is not of the highest intensity. There exist multiple lower energy peaks, among which the peak $\sim$14$-$20 meV below the FX is of the highest intensity. Upon careful observation of multiple samples, the highest intensity peak is found to consist of one, two or three closely spaced peaks. Fig~\ref{fig2} (a-c) shows the PL of the samples, which exhibits a single, double (RX$'$ and RX$''$) and triple (RX$'$, RX$''$ and RX$'''$) split in the highest intensity emission peak.  These single, double, and triple  emission features were observed in multiple samples.  Fig~\ref{fig2} (d-f) shows typical single, double (with energy separation $\Delta_1$) and triple (with energy separations $\Delta_2$ and $\Delta_3$)  peak-split of the  peaks in different samples. The distribution of these splittings $\Delta_1, \Delta_2 \ \rm{and} \ \Delta_3$  across samples showing these values lies in the range of $\sim$1-2 meV (Fig~\ref{fig2} (g-i)). This signature peak-split has previously been associated with Rashba exciton in various \ch{CsPbX3} crystals \cite{becker2018bright,cho2022exciton,sercel2019exciton,Tamarat_2020, Fu_2017}. We assign these peaks to originate from Rashba spin-orbit split triplet \cite{yao2025crystal,wu2019indirect,becker2018bright,sercel2019exciton} and define the mean position of the splitted peaks as Rashba exciton (RX). We estimate the Rashba coupling strength parameter $\alpha$  from the expression:   
\begin{equation}
 \alpha^2 =  \frac{2\hbar^2  (E_{FX}-E_{RX})}{m^*}  
\end{equation}

where,   $m^*$ is the effective mass of exciton (0.126 m$_0$)\cite{yang2017impact}. The estimated values of $\alpha$ $\sim$1.39, close to the value previously reported\cite{yao2025crystal}. Apart from the Rashba peaks, multiple distinct peaks were noticed in lower energy sides; these peaks are likely to involve trions, biexcitons, defect-bound excitons, or phonon replicas\cite{wu2019indirect}.

Fig~\ref{fig2} (j) shows a colour map of PL of the sample showing double split Rashba peaks at different powers at 77 K, the peaks showing no  energy shift with varying power.  Fig~\ref{fig2} (k) shows the normalised PL at different powers, where we observed no changes in the position and relative intensity. These observations suggest the absence of  emission from defect states, and the lower energy peak intensities scale as the highest intensity peak, hinting towards the same excitonic origin of those peaks\cite{lin2024strong}. It is important to note that metal halide perovskites, including \ch{CsPbBr3}, are known to have defect tolerance\cite{kang2017high}. 

\begin{figure*}[t!]
  	\includegraphics[width=0.7\linewidth]{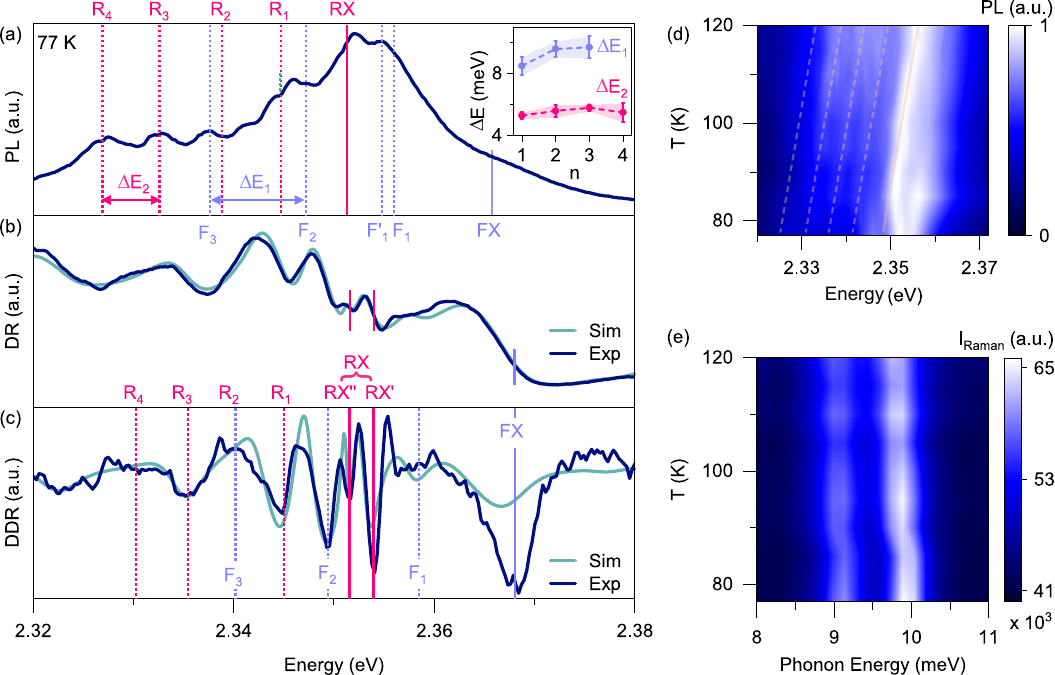}
  	\caption{ \label{fig3} Low-temperature (77 K) correlated PL, reflectance, and Raman spectra. (a) PL spectra showing equally separated phonon replica peaks of the FX and RX, with an  additional peak F$_1'$. The inset shows the replica number (n) vs. $\Delta \rm{E}$ plot for the FX and the RX replicas. (b) Differential reflectance (DR) spectrum. The teal curve shows the simulated DR spectra. (c) First derivative of the differential reflectance (DDR) spectrum. The dips in the DDR correspond to a transition energy, where equally separated phonon replicas of FX and RX are observed. (d) Temperature-dependent PL energy separation, showing a constant energy separation across the given temperature range. (e) Temperature-dependent Raman spectra, exhibiting two strong peaks around 9 meV and 10 meV, exhibit nearly constant intensity and energy within the given temperature regime.} 
  \end{figure*}

To study the origin of the emissions other than FX and RX, we further analysed the PL and reflectance contrast spectra.   Fig~\ref{fig3} (a) shows the PL spectrum. In addition to the FX and RX, the other distinct peaks are found, and we noticed that all these peaks, including the FX and RX, can be grouped into two distinct sets of peaks, with the highest energy peak of the set being the FX and RX. The peaks within the sets exhibit an equally spaced characteristic. These equally separated peaks can be assigned as phonon replicas of FX and RX (SI S3).
 
First, we will discuss the replicas of FX. In Fig~\ref{fig3} (a), the peaks indicated by broken blue lines, FX, F$_{1}$, F$_{2}$ and F$_{3}$ exhibit an approximately equal spacing of $\Delta\rm{E}_1$ $\sim$9-10 meV. We assign these peaks to phonon replicas of the first exciton. Additionally, another peak (F$'_{1}$) was observed, separated by around 11 meV below the FX. S3 Table 1 in the Supplementary Information summarises the observed peaks, their energies, and their energy differences relative to the FX.

 The Rashba peak exhibits two main peaks: 2.352 eV and 2.350 eV. Interestingly, these two primary Rashba peaks also display phonon replicas with equal energy spacing, similar to what is observed for FX. To simplify our discussions ahead, we have assigned the mean position of the   RX replica doublets as  R$_1$, R$_2$,  R$_3$, and R$_4$.  
The energy separation of these replicas with respect to the mean of the main Rashba peak RX is indicated by the purple lines in Fig~\ref{fig3} (a) is  $\Delta\rm{E}_2$. It is found that $\Delta\rm{E}_2\sim$5-6 meV. S3 Table 2 in Supplementary Information provides a summary of these peaks, their energies, and their energy differences relative to the Rashba exciton position RX. The inset in Fig.~\ref{fig3} (a) shows the variation of $\Delta \rm{E} = (\rm{E}_{\rm{ZPL}}-\rm{E}_{\rm{n}})/n$ with the replica number(n), where $\rm{E}_{\rm{ZPL}}$ and $\rm{E}_{\rm{n}}$ corresponds to the energy of the zero phonon line (ZPL), i.e. FX(RX) and the `n'-th phonon replica (n $=$ 1, 2, 3, etc.) of FX(RX). This summarises the calculated energy separations for all the replicas. For the `n'-th phonon replica of FX and RX, $\Delta $E  is found to be nearly constant and with mean $\Delta\rm{E}_1 \sim$9.3$\pm$0.4 meV and $\Delta\rm{E}_2 \sim$5.6$\pm$0.2 meV.

Fig~\ref{fig3} (b) (blue curve) displays the differential reflectance (DR) spectrum calculated from the reflected signal from the sample (R$_{\rm{sample}}$) and the substrate (R$_{\rm{substrate}}$) as,
\begin{equation}
   DR = (\rm{R}_{\rm{sample}} -\rm{R}_{\rm{subtrate}})/ \rm{R}_{\rm{subtrate}}
\end{equation}

For a clearer picture, the Differential Reflectance (DR) was simulated using a multi-Lorentz oscillator model, which agrees well with experimental data\cite{baranowski2019phase}. The simulated curves, shown as a green curve in Fig~\ref{fig3} (b), are in good agreement with the experimental data.  To better identify transition levels, the First Derivative of Reflectance (DDR) was plotted as shown in Fig~\ref{fig3} (c) (blue curve). Each dip in the DDR corresponds to a transition energy, blueshifted by $\sim$1$-$2 meV compared to PL peaks. The FX transition level  and its three phonon replicas are clearly seen with an energy separation $\Delta\rm{E}_1$ of $\sim$9 meV, matching PL data. The Rashba split peaks (RX$'$ and RX$''$) are distinct minima with a $\sim$2 meV separation, strengthening their assignment. RX replicas are also observed, separated by $\sim$5 meV, with the second overlapping the FX's third replica, similar to PL.

The temperature dependence of the PL and Raman spectra (Fig~\ref{fig3} (d,e)) provides insight into the phonon replicas. PL peak energy positions (Fig~\ref{fig3}  (d))  maintain an approximately equal separation up to 120 K. The Raman spectra (Fig~\ref{fig3} (e) show stable strong peaks around 9 meV and 10 meV, with their positions largely unchanged across this temperature range. This signature for these phonon modes, as well as for the lower energy phonon modes, is also previously reported in this temperature window\cite{hoffman2023understanding,cohen2022diverging}. This stability in both the PL energy separation and Raman peak positions  supports the assignment of these features as exciton-phonon replicas\cite{lin2024strong,du2017temperature,le2024wannier}.

\begin{figure*}[t!]
	\includegraphics[width=0.7\linewidth]{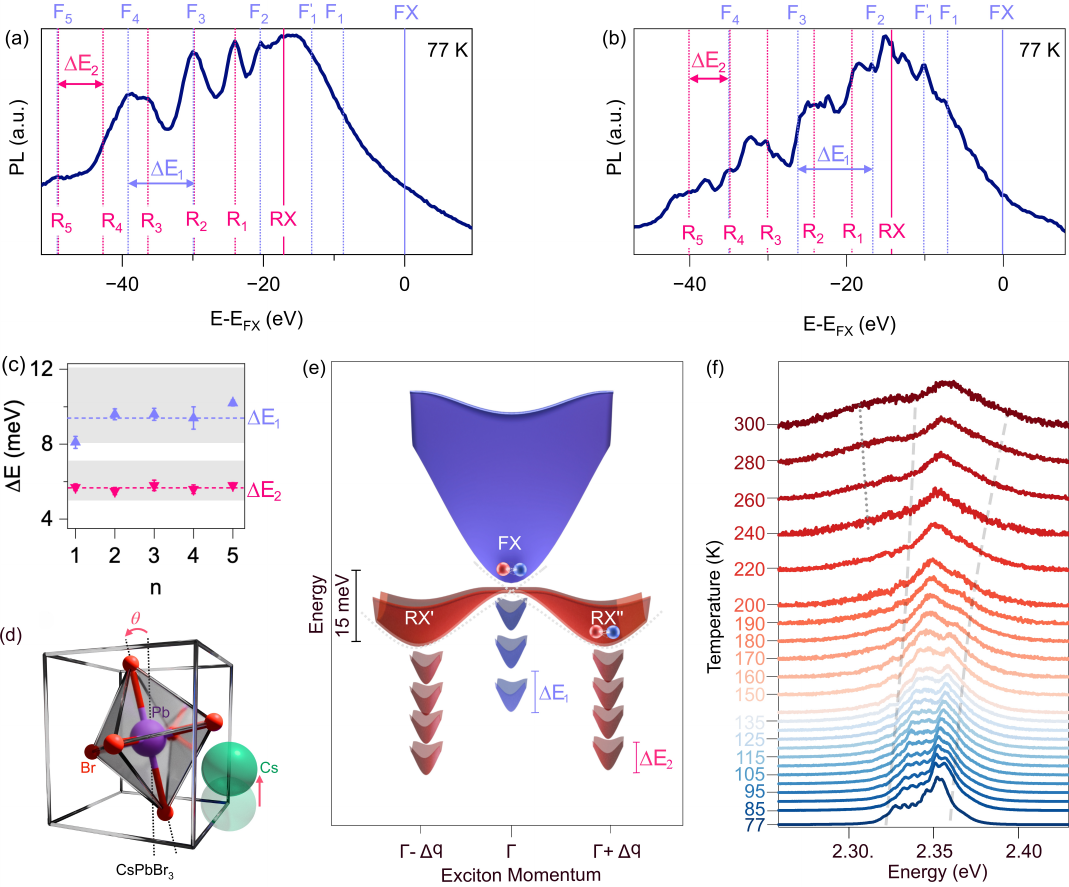}
	\caption{ \label{fig4} Structural and recombination pathways, with samples comparison.  (a) and (b) show normalised PL spectra of samples showing single split and triple split Rashba peaks, as a function of the energy separation of the RX and all replicas with respect to FX, respectively. (c) n vs mean $\Delta\rm{E}$ for FX and RX of the samples. The dotted lines indicate the mean phonon energy involved in the replica generation process ($\Delta\rm{E}_1$ $\sim$9.4 meV and  $\Delta\rm{E}_2$ $\sim$5.6 meV). The shaded area indicates the available phonon energy bands.  (d) Crystal structure of \ch{CsPbBr3}. The slight displacement of the Cs$^+$ cation within the unit cell leads to its occupation at two distinct positions. The Pb-Br-Pb bond also tilts by an angle $\theta$ due to this displacement. (e) Schematic illustration of the various exciton levels, including the FX, RX, and their respective phonon replicas. The discrete replica states are depicted lying under their corresponding energy valleys. (f) Temperature evolution of PL spectra of the sample showing a double-split Rashba peak. With increasing temperature, sharp phonon replica features broaden and merge, evolving into a single peak with a hump at room temperature (enclosed by dashed grey lines). The dotted grey line (240-300 K) shows the emergence of the high-temperature phonon replica region.}
\end{figure*}

Fig~\ref{fig4} (a,b) show the normalised PL spectra of samples showing single split and triple split Rashba peaks, respectively, as a function of the energy separation from $\rm{E}_{\rm{FX}}$.  The single split case shows five FX and five RX phonon replicas, with some overlaps observed (e.g., F$_3$ with R$_2$). The triple split case shows four FX and five RX replicas, also with overlaps attributed to accidental energy matching. In both cases, the phonon replica separations are consistent: $\Delta\rm{E}_1$ is $\sim$9-10 meV and $\Delta\rm{E}_2$ is $\sim$5-6 meV. Since these separations remain unchanged across samples of different thicknesses, an interference effect is unlikely. Fig~\ref{fig4} (c) indicates the mean $\Delta\rm{E}_1$ and $\Delta\rm{E}_2$ for all the samples. The average phonon energies, $\Delta\rm{E}_1 \sim$9.4$\pm$0.2 meV and $\Delta\rm{E}_2 \sim$5.6$\pm$0.1 meV, shown in the shaded grey area are close to the known phonon bands\cite{hoffman2023understanding,guo2019dynamic,gao2021metal,iaru2021frohlich}.

The Huang-Rhys(HR) factor is the standard observable for quantifying the exciton-phonon coupling strength. This factor quantifies the intensity of phonon side bands relative to the main zero-phonon line \cite{zhu2024quantifying}. 
Single phonon HR factor ($S^1_\nu$) for a phonon mode with energy $\nu$ meV is estimated for FX and RX \cite{theoretica2025}. For FX $S^1_9$ ranges from 1.6 to 3.6 and for RX  $S^1_6$ ranges from 0.75 to 0.94 in the samples studied. This results in a clear contrast with the reported decrease of the HR factor with increasing crystal size \cite{zhu2024quantifying,cho2022exciton}. Notably, the FX $S^1_9$ is, to the best of our knowledge, the highest reported single phonon HR factor in \ch{CsPbBr3} microcrystals. Pronounced exciton-phonon coupling has primarily been observed in nanocrystals of perovskites at exceedingly low temperatures, its strength tending to diminish in larger crystals \cite{zhu2024quantifying,amara2023spectral,cho2022exciton}. However, our observations suggest the presence of this coupling in microcrystals, which is quite uncommon; this invites reexamination of the established view that particle size dictates exciton-phonon behaviour in lead halide perovskites.  We have also noticed the existence of $\rm{F}_1'$ approximately below 11 meV from the FX for these two samples.

Our observations confirm two key features: the existence of a low-energy Rashba exciton and distinct phonon replicas for both high-energy and Rashba excitons in \ch{CsPbBr3} microcrystals. To explore the microscopic origins, we examine the crystal structure as shown Fig~\ref{fig4} (d). Bending of the Pb-Br-Pb bond with a subtle displacement of the Cs$^+$ cation within the unit cell is reported in \ch{CsPbBr3} nanocrystals\cite{li2020evidence}. This displacement of the $\text{Cs}^+$ cation within the orthorhombic unit cell breaks inversion symmetry\cite{anilkumar2024near}, which can induce a strong Rashba spin-orbit interaction. Additionally, octahedral polar fluctuations or surface termination can cause the dynamic structural distortion necessary for the Rashba effect\cite{wu2019indirect,yaffe2017local, niesner2016giant,anandan2023spin,wang2025octahedral,krach2023emergence,yao2025crystal}.

Phonon replicas manifest as distinct signatures in both DR and PL spectra\cite{dyksik2024polaron,baranowski2019phase,posmyk2024bright}, granting us an opportunity to explore exciton-phonon coupling in this system. In this study, we observed phonon replicas as equally separated lines with energy separations of $\sim$6 meV and $\sim$9 meV. As we previously discussed, Raman spectra revealed prominent phonon modes within the energy band between 8-12 meV at 77 K. Alongside this dominant phonon mode, other lower-energy phonon modes, specifically between 5-7 meV, are observed in \ch{CsPbBr3}\cite{cho2022exciton,hoffman2023understanding,guo2019dynamic,gao2021metal,iaru2021frohlich,yaffe2017local}. In our study, the phonon modes within the 8-12 meV band are associated with the high energy exciton, leading to the prominent high energy exciton phonon replicas that we observe in both DR and PL spectra. The lower-energy phonon modes in the 5-7 meV band remain crucial in the formation of Rashba exciton replicas. These phonon modes are related to the bending of the Pb-Br-Pb bonds\cite{cho2021luminescence,cho2022exciton,miyata2017large,yamada2022electron,lanigan2021two}. As we previously discussed, a displacement of the Cs$^+$ cation with the Pb-Br-Pb bond bending containing heavy atoms like Pb gives rise to a strong Rashba spin-orbit coupling effect in \ch{CsPbBr3}.  Thus, these low-frequency modes, through local structural distortions, can create a deformation potential which may induce exciton-phonon interaction,\cite{lahnsteiner2024tuning,ma2020local,yamada2022electron} leading towards phonon replicas in the DR and emission spectra. Fig~\ref{fig4} (e) represents a schematic summary of the proposed model for low-temperature phonon replicas of FX and RX. The $\rm{F}_1'$ is probably due to the coupling of the 11 meV phonon mode with FX.

To understand temperature's effect on exciton interactions, we analyzed the evolution of PL spectra from 77 K to 300 K. As temperature rises, thermal broadening causes the distinct low energy phonon replicas to merge into a single, broad peak around 150 K\cite{lin2024strong, StevenLouie2022}. This peak further merges with the Rashba exciton   peak near 190 K, forming a dominant peak that persists up to 300 K (enclosed by dashed grey lines). Interestingly, an additional, lower energy broad peak emerges at 240 K, and its relative intensity increases with temperature (indicated by dotted grey lines). At high temperatures, the PL spectrum resolves into multiple peaks (see SI S5), with the Rashba splitting ($\sim$24 meV) increasing \cite{wu2019indirect}. As lower energy Raman modes vanish\cite{hoffman2023understanding},(see SI S6) a prominent $\sim$58 $\pm$1.3  meV phonon replica separation emerges near room temperature, attributed to the combination of the 19 meV and 38 meV Raman active modes\cite{ma2020local}. These findings demonstrate a temperature driven crossover in phonon replicas, from low energy phonon dominance at low temperatures to  LO phonon dominance at higher temperatures, necessitating further theoretical investigation into lattice dynamics, anharmonicity, and screening effects\cite{theoretica2025}.

\begin{figure}[t!]
\centering
	\includegraphics[width=0.7\linewidth]{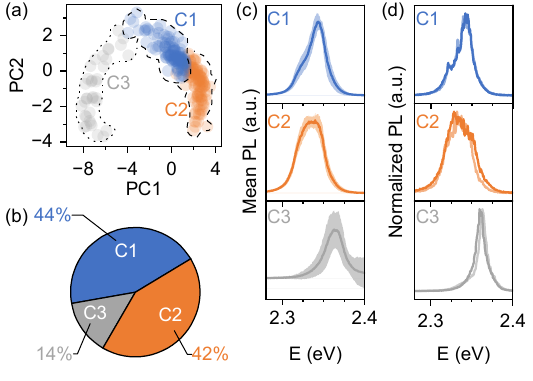}
	\caption{ \label{fig5}Statistical study of the emission characteristics for 260 samples. (a) Principal component scatter plot and k-means clustering of the emission lineshapes. The emission lineshape is classified into three clusters: C1 (blue), C2 (orange) and C3 (grey). C1 and C2 show the highest density in the scattered plot, referring to the most probable lineshapes. (b) Percentage distribution of the clusters. (c) Mean emission curve of clusters C1 (top panel), C2 (middle panel) and C3 (bottom panel) (in solid line). The highlighted region in each panel shows the variation of the spectra within the cluster across the mean spectra. (d) Two representative spectra from clusters C1 (top panel), C2 (middle panel) and C3 (bottom panel).}
\end{figure}

To better understand the variations in the spectral signatures, we performed a statistical analysis on low-temperature (77 K) normalised PL spectra from 260 samples using principal component analysis (PCA) and k-means clustering. The k-means analysis identified three distinct clusters: C1 (blue), C2 (orange), and C3 (grey), as shown in Fig~\ref{fig5} (a) (SI S7). Most samples fall into clusters C1 or C2, with only a small number of samples exhibiting the emission signatures of cluster C3. This is evident from the high-density regions of C1 and C2 in Fig~\ref{fig5}(a) and the percentage distribution for each cluster shown in Fig~\ref{fig5} (b). The mean spectrum for each cluster is shown in Fig~\ref{fig5} (c), where highlighted regions indicate the standard deviation. Notably, the mean spectra for C1 and C2 show low intra-cluster variation. In contrast, the mean spectrum for C3, centred at a higher energy, exhibits comparatively high variation. Therefore, C3 represents a cluster of outliers with higher-energy emissions. To illustrate the nature of the emission, Fig~\ref{fig5} (d) presents two representative spectra from each cluster. The spectra in C1 and C2 prominently feature FX, RX, and their phonon replicas, and the relative strength of these replicas distinguishes the two clusters. In C1, the phonon replica region is lower in intensity, whereas in C2, the replicas are significantly stronger, with numerous closely spaced peaks merging to form a broad emission band (SI  Figure S7e ). In conclusion, most of the samples studied show emission dominated by FX, RX, and their respective phonon replicas.

In summary, our comprehensive study on \ch{CsPbBr3} microcrystals reveals a rich, complex optical response, demonstrating the unambiguous observation of coexisting high-energy and Rashba excitons, each exhibiting a distinct phonon replica series that highlights multiple exciton–phonon coupling channels and challenges established views on particle-size dependence. Furthermore, the coupling is dynamic, showing a progressive merging of replicas with rising temperature and a crossover to a dominant longitudinal optical (LO) phonon coupling regime at room temperature, demonstrating adaptability in their interaction pathways. The intrinsic, exciton selective  coupling enables a materials-by-design paradigm for quantum  control using perovskites.

HLP acknowledges funding support from the National Mission on Interdisciplinary Cyber-Physical Systems (NM-ICPS) of the DST, Government of India, through the I-HUB Quantum Technology Foundation, Pune, India. SC thanks the Prime Minister’s Research Fellowship, Govt. of India, for providing a research fellowship. AR acknowledges funding support from DST SERB grant no. CRG/2021/005659. AS acknowledges funding from Department of Science and Technology Nanomission CONCEPT grant (NM/TUE/QM-10/2019). AA acknowledges financial support from the following projects funded by the Government of India: NM-ICPS of the DST through the I-HUB Quantum Technology Foundation (Pune, India), Project No. CRG/2022/007008 of SERB, MoE-STARS project No. MoE-STARS/STARS-2/2023-0912, CEFIPRA CSRP Project No. 7104-2, and DST National Quantum Mission project No. DST/QTC/NQM/QMD/2024/4 (G).CRG/2021/005659.

\bibliography{sn-bibliography}

\end{document}